\documentclass[onecolumn,notitlepage,letterpaper,superscriptaddress,nofootinbib,longbibliography]{revtex4-2}

\usepackage{amsmath}
\usepackage{graphicx}
\usepackage{amsfonts}
\usepackage{latexsym}
\usepackage{bbold}
\usepackage{calligra}
\usepackage{float}
\usepackage{ulem}
\usepackage{inputenc}
\usepackage{subfig}
\usepackage{xspace}
\usepackage{url}
\usepackage{epstopdf}
\usepackage{tikz}
\usepackage{amsthm}
\usepackage{cancel}
\usepackage{comment}
\usepackage{array}
\usepackage{subfig}
\usepackage{color}
\usepackage{tensor}
\usepackage{xcolor}
\usepackage{orcidlink}
\usepackage{comment}
\usepackage{physics}
\usepackage[mathscr]{eucal}
\usepackage{multirow}
\usepackage{graphicx}
\usepackage{amssymb}
\usepackage{amsmath}
\usepackage{subfig}
\usepackage{xcolor}

\usepackage{hyperref}

\hypersetup{
    colorlinks=true,
    linkcolor=blue,
    citecolor=blue,
    filecolor=magenta,
    urlcolor=blue,
    pdfstartview=FitH,
    pdfmenubar=true,
    pdftoolbar=true
}

\newcommand{\be}{\begin{equation}}
\newcommand{\ee}{\end{equation}}
\newcommand{\beq} {\begin{equation}}
\newcommand{\eeq} {\end{equation}}
\newcommand{\ba}{\begin{eqnarray}}
\newcommand{\ea}{\end{eqnarray}}

\begin{document}

	\title{Inflation with Nieh-Yan-like terms in metric-affine gravity}
	
	\author{Ilaria Andrei}
    \email{ilaria.andrei@ut.ee}
    \affiliation{Laboratory of Theoretical Physics, Institute of Physics, University of Tartu, W. Ostwaldi 1, 50411 Tartu, Estonia.}
    
    \author{Christian Dioguardi}
    \email{christian.dioguardi@kbfi.ee}
    \affiliation{Laboratory of High Energy and Computational Physics,
    National Institute of Chemical Physics and Biophysics, R{\"a}vala pst.~10, Tallinn, 10143, Estonia}

    \author{Damianos Iosifidis}
    \email{d.iosifidis@ssmeridionale.it}
    \affiliation{Scuola Superiore Meridionale, Largo San Marcellino 10, 80138 Napoli, Italy and INFN– Sezione di Napoli, 
    Via Cintia, 80126 Napoli, Italy}
    
    \author{Laur Järv}
    \email{laur.jarv@ut.ee}
    \affiliation{Laboratory of Theoretical Physics, Institute of Physics, University of Tartu, W. Ostwaldi 1, 50411 Tartu, Estonia.}
    
    \author{Antonio Racioppi}
    \email{antonio.racioppi@kbfi.ee}
    \affiliation{Laboratory of High Energy and Computational Physics,
    National Institute of Chemical Physics and Biophysics, R{\"a}vala pst.~10, Tallinn, 10143, Estonia}

    \author{Margus Saal}
    \email{margus.saal@ut.ee}
    \affiliation{Laboratory of Theoretical Physics, Institute of Physics, University of Tartu, W. Ostwaldi 1, 50411 Tartu, Estonia.}

	\date{\today}
\begin{abstract}
We study single-field slow-roll inflation in metric-affine gravity with a scalar field non-minimally coupled to the non-Riemannian Ricci scalar and to the divergences of the torsion and nonmetricity vectors, a structure that generalizes the well-known Nieh-Yan term. By imposing projective coherence of the matter sector and solving the connection field equations, we integrate out torsion and nonmetricity and obtain an equivalent Einstein-frame formulation in which the metric-affine couplings are encoded in a modified kinetic function and potential. For the choice of coupling functions $\mathcal{A}(\phi) = M_P^2 + \xi \phi^2$ to the non-Riemannian Ricci scalar, $\mathcal{C}_i(\phi) = \xi_i \phi$ to the Nieh-Yan-like terms and a monomial Jordan-frame potential $\mathcal{V} \propto \phi^k$, we show that in the limit of a large positive effective Nieh-Yan-like coupling $\bar{\xi}$ the canonical field satisfies $\chi \sim \phi^2$, the Jordan-frame field values during inflation become sub-Planckian, and the Einstein-frame potential reduces to $U \sim \chi^{k/2}$. We compute the slow-roll predictions numerically for quartic and quadratic Jordan-frame potentials and compare them with the current CMB constraints from Planck, BICEP/Keck, ACT, and SPT. We find that intermediate values of $\bar{\xi}$ can restore the compatibility of non-minimally coupled Palatini inflation with observations: in the quartic case, the model predicts a tensor-to-scalar ratio within reach of next-generation CMB experiments for $\bar{\xi}\lesssim10^4$, while in the quadratic case the coupling cures the $\eta$-problem arising for $\xi \gtrsim 10^{-2}$ and yields viable predictions for $10^{-2}\lesssim\bar{\xi}\lesssim 10^2$. In the negative $\bar{\xi}$ regime, the model does not improve upon standard Palatini inflation, though it can still produce distinct, testable predictions.
\end{abstract}

	\maketitle
	
	\allowdisplaybreaks
\section{Introduction}

The $\Lambda$CDM model, the standard cosmological model, is based on the cosmological principle, which assumes that the Universe is homogeneous and isotropic on sufficiently large scales. Observations of the cosmic microwave background (CMB) \cite{Planck:2018vyg} strongly support such an assumption, showing that the Universe is astonishingly uniform across hundreds of megaparsecs. Nevertheless, explaining such uniformity, together with the approximately flat geometry of the Universe and the absence of exotic relics, requires an additional mechanism acting in the very early Universe. The most compelling solution is a brief phase of rapid accelerated expansion~\cite{Starobinsky:1980te, Guth:1980zm, Linde:1981mu, Albrecht:1982wi}, commonly referred to as inflation. Such an exponential expansion not only solves the aforementioned issues but also provides a natural origin for the tiny primordial fluctuations that later evolve into galaxies, clusters, and the cosmic web. 

The simplest realization of inflation is through a scalar field, the inflaton, whose potential energy density drives the accelerated expansion. The specific shape of the inflaton potential determines both the duration of inflation as well as the characteristics of the resulting primordial perturbations like the scalar spectral index $n_s$ and tensor-to-scalar ratio $r$, see, e.g., \cite{Martin:2013tda}. Recent observations of the CMB by Planck, BICEP/Keck, ACT, and SPT \cite{Planck:2018jri, BICEPKeck:2021sbt, AtacamaCosmologyTelescope:2025nti, SPT-3G:2025bzu} have ruled out the simplest minimally coupled models, motivating the exploration of more elaborate constructions. In particular, the latest combination from the BICEP/Keck collaboration~\cite{BICEP:2021xfz} points to the Starobinsky model \cite{Starobinsky:1980te} and Higgs-inflation \cite{Bezrukov:2007ep} as representatives of the most favoured class of models. These can be described by a scalar field non-minimally coupled to the Riemannian curvature scalar in the action (e.g., \cite{Galante:2014ifa, Jarv:2016sow} and references therein). Although the inclusion of recent ACT data shifts the preferred region towards higher values of $n_s$ and puts these models under tension with observations \cite{AtacamaCosmologyTelescope:2025nti}, taking the radiative corrections into account can restore their compatibility with the ACT constraints \cite{Gialamas:2025kef, Wolf:2025ecy, Ellis:2025bzi, Yuennan:2025inm}. In this context, it is also important to note that, as explained in detail in \cite{Ferreira:2025lrd, Balkenhol:2025wms}, the shift towards higher $n_s$ values is due to unresolved tension between CMB and baryon acoustic oscillations (BAO) data. Therefore, in order to be as cautious as possible, we will consider constraints with and without BAO data.

When the data are read to favour scalar fields with non-minimal couplings to gravity, the choice of geometric framework becomes important. General relativity, in the original setup, assumes all gravitational degrees of freedom to be encoded in the metric; the Levi-Civita connection is fixed by it. On the other hand, in the Palatini formulation, the metric and the (symmetric) connection are taken to be independent, while the usual Einstein field equations arise only after solving the connection field equations. However, the equivalence of these two approaches is broken once non-minimal couplings are introduced to the action, and the choice of the geometric framework plays a critical role in determining the inflationary dynamics and predictions \cite{Bauer:2008zj, Kallosh:2022feu, Marzola:2016xgb, Jarv:2017azx, Racioppi:2018zoy, Enckell:2018hmo, Jarv:2020qqm, Gialamas:2020snr, Dioguardi:2021fmr, Dioguardi:2022oqu, Dioguardi:2023jwa, Jarv:2024krk}. 

An even broader framework is provided by metric-affine gravity (MAG) \cite{Hehl:1976my, Hehl:1976kj, Hehl:1994ue}. In this setting, the metric independent part of the affine connection is characterised by nonmetricity and torsion (antisymmetric part) contributions. Therefore, a generic gravitational action in MAG may include various combinations of contracted curvature, torsion, and nonmetricity tensors, or functions thereof. Equally importantly, MAG theories allow matter fields to couple to the independent connection, this gives rise to hypermomentum. The hypermomentum tensor is formally defined as the variation of the matter part of the action with respect to the connection \cite{hehl1976hypermomentum, hehl1978hypermomentum}, and it can be considered as a generalisation of the concept of the spin tensor by including also the dilation (trace) and shear (symmetric traceless) parts. The hypermomentum and the usual energy-momentum tensor constitute the sources of the MAG field equations. While the theoretical options of coupling spinor and vector fields to the affine connection are rather restricted by consistency considerations \cite{BeltranJimenez:2020sih, Rigouzzo:2023sbb}, scalar fields are subject to fewer such restrictions and can generate hypermomentum in a number of forms once one goes beyond minimal coupling.

Inflationary models in MAG may be divided into two principal classes. First, in MAG theories where the connection components propagate as dynamical fields, suitable torsion or nonmetricity modes may serve to run inflation, see e.g. \cite{Salvio:2022suk, Karananas:2025xcv, Salvio:2025izr}. However, it can be precarious to construct models of this type, because dynamical connection is often plagued by ghost degrees of freedom \cite{BeltranJimenez:2019acz, Percacci:2020ddy, Barker:2025xzd}. Second, in MAG theories with nondynamical connection, torsion and nonmetricity can be algebraically eliminated from the equations. But if nontrivial hypermomentum is present, the constraint on torsion and nonmetricity will be nontrivial as well, and hence the form of the remaining equations will bear the mark. Thus, introducing a coupling between the scalar field and affine connection into the action will reshape the effective inflaton sector, as the resulting effective metric theory will contain a modified scalar field kinetic term and potential, providing a useful mechanism to tune inflation \cite{Shimada:2018lnm,Shaposhnikov:2020gts,Shaposhnikov:2020frq,Mikura:2020qhc,Mikura:2021ldx,Rigouzzo:2022yan,Gialamas:2022xtt,DiMarco:2023ncs,Gialamas:2024jeb,Capozziello:2024lsz,Gialamas:2024iyu,Racioppi:2024pno,Gialamas:2024uar,He:2025bli,Iosifidis:2025wrv,Katsoulas:2025srh,Racioppi:2025igu,Dimopoulos:2026iwq,Kraiko:2026nas,Racioppi:2024nmi,Hassan:2026ddx}. Similar mechanisms have also been investigated in the context of late-time cosmology; see, e.g., \cite{Iosifidis:2020gth, Iosifidis:2021nra, Andrei:2024vvy, Andrei:2026ryg}.

In this paper, we consider single-field slow-roll inflation in a metric-affine framework, with a scalar field non-minimally coupled to the non-Riemannian curvature scalar, and derivative couplings between the scalar field and the possible self-contractions of torsion and nonmetricity tensors, motivated by an effective field theory (EFT) approach. These couplings generalize the standard Nieh-Yan term which has been already extensively studied in the context of inflation \cite{Langvik:2020nrs, Shaposhnikov:2020gts, Piani:2022gon, He:2024wqv}. 
The paper is organized as follows. In section \ref{sec:setup} we introduce the metric-affine framework, highlighting the differences with the standard metric formulation of gravity. In section \ref{sec:action}, we introduce our starting action and derive the Einstein-frame equivalent formulation of our model. In section \ref{sec:sr}, we briefly introduce the slow-roll formalism to compute the predictions of the model for inflation. In section  \ref{sec:model}, we use the slow-roll approximation to compute the CMB observables and compare the predictions of the model with the current datasets coming from cosmological observations. Finally, in section \ref{sec:conclusion} we summarize the main results of this work and draw our conclusions.

\section{Setup} \label{sec:setup}

In what follows, we work in a 4-dimensional non-Riemannian manifold with metric signature $(-,+,+,+)$ and consider natural units $\hbar=c=1$. In the metric-affine approach, the fundamental geometric variables are the metric $g_{\mu\nu}$, with associated Levi-Civita connection $\tilde{\nabla}$, with coefficients the Christoffel symbols:
\begin{equation}
    \label{lcconn}
    \tilde{\Gamma}^\lambda_{\phantom{\lambda}\mu\nu} = \frac{1}{2} g^{\rho\lambda}
    \left(\partial_\mu g_{\nu\rho} + \partial_\nu g_{\rho\mu} - \partial_\rho g_{\mu\nu}\right),
\end{equation}
and an independent affine connection $\nabla$ with coefficients $\Gamma^{\lambda}{}_{\mu \nu}$ that can be written as
\begin{equation}\label{eq:affine}
   \Gamma^{\lambda}{}_{\mu \nu} = \tilde{\Gamma}^{\lambda}{}_{\mu \nu} + N^{\lambda}{}_{\mu \nu}\,,  
\end{equation}
where the tensor $N^\lambda{}_{\mu\nu}$ is called distortion. The distortion tensor can be expressed in terms of two geometrical quantities, nonmetricity $Q$ and torsion $S$.
The nonmetricity tensor is defined as
\beq\label{eq:nonmetricity}
    Q_{\alpha\mu\nu}:=-\nabla_{\alpha}g_{\mu\nu}  = -\partial_{\alpha}g_{\mu\nu} + \Gamma^{\lambda}_{\phantom{\lambda}\mu\alpha}g_{\lambda\nu}
    +\Gamma^{\lambda}_{\phantom{\lambda} \nu\alpha}g_{\lambda\mu},
\eeq
and the torsion tensor is defined by
\beq \label{eq:torsion}
S_{\mu\nu}^{\phantom{\mu\nu}\lambda}:=\Gamma^{\lambda}{}_{[\mu\nu]} \,,
\eeq
where the square brackets denote antisymmetrization, $2 A_{[\mu\nu]}\equiv A_{\mu\nu} - A_{\nu\mu}$. Using \eqref{eq:affine}-\eqref{eq:torsion}, we can explicitly write the relation between distortion, nonmetricity, and torsion as:
\beq
\label{distortion}
    N^\lambda{}_{\mu\nu} = {\frac12 g^{\rho\lambda}\left(Q_{\mu\nu\rho} + Q_{\nu\rho\mu}
    - Q_{\rho\mu\nu}\right)} - {g^{\rho\lambda}\left(S_{\rho\mu\nu} +
    S_{\rho\nu\mu} - S_{\mu\nu\rho}\right)}\,.
\eeq
From torsion and nonmetricity, we can also construct the following vector contractions
\begin{equation}
\label{eq:QSvectors}
    t^{\alpha } := \epsilon^{\alpha \beta \gamma \rho} \,S_{ \beta \gamma \rho}  \hspace{1 cm}    
    S_{\alpha } := S_{\alpha \beta }{}^{\beta } \hspace{1 cm}
    Q_{\mu }:= g^{\alpha \lambda } Q_{\mu \alpha \lambda }  \hspace{1 cm} 
    q_{\mu } := g^{\alpha \lambda } Q_{\alpha \lambda \mu } \,,
\end{equation}
where $\epsilon_{\alpha \beta \gamma \rho}$ is the completely antisymmetric Levi-Civita tensor.
The non-Riemannian curvature tensor associated with the affine connection is defined by
\begin{equation}\label{Riemanntensor}
R^\mu{}_{\nu\alpha\beta} := \partial_\alpha \Gamma^\mu{}_{\nu\beta} 
- \partial_\beta \Gamma^\mu{}_{\nu\alpha} 
+ \Gamma^\mu{}_{\lambda\alpha}\Gamma^\lambda{}_{\nu\beta} 
- \Gamma^\mu{}_{\lambda\beta}\Gamma^\lambda{}_{\nu\alpha},
\end{equation}
from which we can construct three Ricci-like tensors:
\beq
    R_{\nu \beta}:=R^{\mu}_{\phantom{\mu} \nu \mu \beta}\,, \qquad 
    \widehat{R}_{\alpha \beta}:= R^{\mu}_{\phantom{\mu} \mu \alpha \beta} \,,\qquad \breve{R}^{\lambda}_{\phantom{\lambda} \kappa}
    :=R^{\lambda}_{\phantom{\lambda} \mu\nu\kappa}g^{\mu\nu} \,.
\eeq
Among these contractions, the only independent scalar is the non-Riemannian Ricci scalar
\beq \label{Ricci scalar}
    R:=R_{\mu\nu}g^{\mu\nu}=-\breve{R}_{\mu\nu}g^{\mu\nu}\,, \qquad  \widehat{R}_{\mu\nu}g^{\mu\nu}=0 \,.
\eeq
The relation between the non-Riemannian $R$ and the Riemannian Ricci scalar $\tilde{R}$ is given by:
\begin{equation}
\label{Rnonriem}
\begin{aligned}
   R &= \tilde{R} +\tfrac{1}{4} Q_{\mu \nu \sigma } Q^{\mu \nu \sigma } -  \tfrac{1}{2} Q_{\mu \nu \sigma } Q^{\nu \mu \sigma } 
   + \tfrac{1}{2} q_{\mu } Q^{\mu } 
   - \tfrac{1}{4} Q_{\mu } Q^{\mu } + 2 \,S_{\mu \sigma }{}^{\nu } S^{\mu }{}_{\nu }{}^{\sigma }\\
   &+ S_{\mu \nu }{}^{\sigma } S^{\mu \nu }{}_{\sigma } + 2 Q_{\mu \nu \sigma } S^{\mu \nu \sigma } + 2  \,q_{\mu } S^{\mu } 
   - 2  \,Q_{\mu } S^{\mu } - 4  \,S_{\mu } S^{\mu } + \tilde{\nabla}_{\lambda }q^{\lambda } 
   -  \tilde{\nabla}_{\lambda }Q^{\lambda } - 4 \, \tilde{\nabla}_{\lambda }S^{\lambda } \,.
   \end{aligned}
\end{equation} 
The non-Riemannian Ricci scalar $R$ in \eqref{Rnonriem} is invariant under the projective transformation \cite{Eisenhart1927,Hehl:1994ue,Sotiriou:2007yd} i.e.
\beq
\label{projective}
    \Gamma^{\lambda}{}_{\mu\nu} \mapsto \Gamma^{\lambda}{}_{\mu\nu}+\delta^{\lambda}_{\mu}\xi_{\nu} \,.
\eeq
A quantity invariant under \eqref{projective} is referred to as a projectively invariant. A special property of this specific transformation, as opposed to a generic vectorial transformation, is that it maps autoparallels $u^{\alpha}\nabla_{\alpha}u^{\mu}=0$ into autoparallels up to a different affine parameter, see \cite{Eisenhart1927}. 

At the level of the connection field equations, projective invariance of the gravitational sector imposes a constraint on the matter sector, namely a vanishing dilation current \cite{Hehl:1994ue} (see the discussion below).

\section{The action}\label{sec:action}

We start from the following action, written in terms of the affine covariant derivative $\nabla$:
\begin{equation}
    \begin{aligned}
       S [g,\Gamma,\phi]
         =&\int \mathrm{d}^{4}x \sqrt{-g}  \,\,\,\Big( \frac{1}{2}\mathcal{A}(\phi)R-\frac{1}{2}\mathcal{B}(\phi) \partial_\mu \phi \partial^\mu \phi - \mathcal{V}(\phi)  -  \mathcal{I}_1(\phi) \nabla_{\mu }Q^{\mu } - \mathcal{I}_2(\phi) \nabla_{\mu }q^{\mu } -  \mathcal{I}_3(\phi) \nabla_{\mu }S^{\mu } -  \mathcal{I}_4(\phi) \nabla_{\mu }t^{\mu }\\
         & \hspace{3 cm}+ \nabla_\mu (\mathcal{I}_1 Q^{\mu }+q^{\mu } \mathcal{I}_2 + \mathcal{I}_3 S^{\mu } + \mathcal{I}_4 t^{\mu }) \Big)\,.
    \end{aligned} \label{eq:S:starting}
\end{equation}
Here $R$ is the non-Riemannian Ricci scalar defined in \eqref{Rnonriem}, $\phi$ is a non-minimally coupled scalar field and $Q^\mu, q^\mu, S^\mu, t^\mu$ are the contractions of nonmetricity and torsion defined in \eqref{eq:QSvectors}.\footnote{For an ordinary scalar inflaton, the coupling to $t^\mu$ is parity odd, as in the Nieh-Yan case, while the derivative couplings to $Q^\mu$, $q^\mu$, and $S^\mu$ are parity even.} The scalar field is non-minimally coupled to the non-Riemannian Ricci scalar through the function $\mathcal{A}(\phi)$, and to the covariant derivatives of torsion and nonmetricity vectors through the functions $\mathcal{I}_i(\phi)$. Starting from \eqref{eq:S:starting}, one can express the affine covariant derivatives into their Levi-Civita parts and distortion contributions. For instance, $\nabla_{\rho}t^{\rho} = N^{\sigma}{}_{\rho \sigma} t^{\rho} + \tilde{\nabla}_{\rho}t^{\rho}$. Analogous terms arise both from the bulk terms and from the total derivative terms, and cancel out. Equivalently, this follows from the identity $(\partial_{\mu}I) A^{\mu}=\nabla_{\mu}(IA^{\mu})-I\nabla_{\mu}A^{\mu}$, where $I$ is a scalar function, which holds independently of the connection used in the covariant derivative. This gives
\begin{equation}
    \begin{aligned}
       S [g,\Gamma,\phi] = &\int \mathrm{d}^{4}x \sqrt{-g}  \,\,\,\Big( \frac{1}{2}\mathcal{A}(\phi)R-\frac{1}{2}\mathcal{B}(\phi) \partial_\mu \phi \partial^\mu \phi - \mathcal{V}(\phi)  -  \mathcal{I}_1(\phi) \tilde{\nabla}_{\mu }Q^{\mu } - \mathcal{I}_2(\phi) \tilde{\nabla}_{\mu }q^{\mu } -  \mathcal{I}_3(\phi) \tilde{\nabla}_{\mu }S^{\mu } -  \mathcal{I}_4(\phi) \tilde{\nabla}_{\mu }t^{\mu } \\
& \hspace{3 cm}+ \tilde{\nabla}_\mu (\mathcal{I}_1 Q^{\mu }+\mathcal{I}_2 q^{\mu } + \mathcal{I}_3 S^{\mu } + \mathcal{I}_4 t^{\mu }) \Big).
    \end{aligned}\label{eq:S:starting1}
\end{equation}
Notice that the $\mathcal{I}_4(\phi) \tilde{\nabla}_{\mu }t^{\mu }$ term is the standard Nieh-Yan term, already studied in \cite{Langvik:2020nrs} in the context of inflation. In this work, we also include the analogous derivative couplings to the other nonmetricity and torsion vectors, that naturally arise in a metric-affine framework. After dropping the boundary term and integrating by parts, we can rewrite the action in a more convenient form:
\begin{equation}
    \begin{aligned}
    S [g,\Gamma,\phi]&=\int \mathrm{d}^{4}x \sqrt{-g} \,\,\,\Big( \frac{1}{2}\mathcal{A}(\phi)R-\frac{1}{2}\mathcal{B}(\phi) \partial_\mu \phi \partial^\mu \phi - \mathcal{V}(\phi) + \, \partial_{\mu}\phi\left(\mathcal{C}_1(\phi) Q^{\mu}+\mathcal{C}_2(\phi) q^{\mu} + \mathcal{C}_3(\phi)  S^{\mu}+\mathcal{C}_{4}(\phi) t^{\mu}\right) \Big),
    \label{action} 
    \end{aligned}
\end{equation}
where we have defined the functions $\mathcal{C}_i(\phi) \equiv \mathcal{I}^{\prime}_i(\phi)$ and the prime denotes differentiation with respect to $\phi$. 

Varying the action \eqref{action} with respect to the affine connection $\Gamma$, we obtain the connection field equations:
\begin{equation}\label{Gfieldeqs}
P_\lambda{}^{\mu\nu} = \frac{1}{M_P^2}\Delta_\lambda{}^{\mu\nu}\,,
\end{equation}
where on the left-hand side we have the Palatini tensor $P_\lambda{}^{\mu\nu}$, defined as the variation of $R$ with respect to the affine connection $\Gamma$,
\begin{equation}\label{palatinite}
    P_\lambda{}^{\mu\nu}:=\frac{\delta R}{\delta \Gamma^{\lambda}{}_{\mu\nu}}= \left( \frac{Q_{\lambda}}{2}+2 S_{\lambda}\right)g^{\mu\nu}
    - (Q_{\lambda}{}^{\mu\nu}+2 S_{\lambda}{}^{\mu\nu})
    +\left( q^{\mu} -\frac{Q^{\mu}}{2}-2 S^{\mu}\right)\delta_{\lambda}^{\nu},
\end{equation}
while the right-hand side contains the hypermomentum tensor. Denoting by $\mathcal{L}$ the full Lagrangian we define
 \begin{equation}\label{hypedef}
     \Delta_{\lambda}^{\phantom{\lambda} \mu\nu} 
   := -\frac{2}{\sqrt{-g}}\frac{\delta ( \sqrt{-g} \,\mathcal{L} )}{\delta \Gamma^{\lambda}_{\phantom{\lambda}\mu\nu}}\Big|_{S_{\alpha\beta\gamma}=Q_{\alpha\beta\gamma}=0} \,.
\end{equation}
For the action \eqref{action}, this gives
\begin{equation}\label{hyper}
    \begin{aligned}
    \Delta_\lambda{}^{\mu\nu} =  & \frac{M_P^2}{ \mathcal{A}(\phi)} \Big( \bigl( g^{\mu \nu } \partial_{\lambda }\phi - \delta_{\lambda }{}^{\nu } \partial^{\mu }\phi\bigr)\, \mathcal{A}^{\prime}(\phi)
    - \bigl( \mathcal{C}_3(\phi)+2\, \mathcal{C}_2(\phi)\bigr) \delta_{\lambda }{}^{\nu } \partial^\mu \phi \\
    &- \bigl(-  \mathcal{C}_3(\phi) + 4 \,\mathcal{C}_1(\phi)\bigr) \delta_{\lambda }{}^{\mu } \partial^\nu \phi - 2\,\mathcal{C}_2(\phi) \,g^{\mu \nu }  \partial_\lambda
    \phi+2\,\mathcal{C}_4(\phi) \,\epsilon_{\lambda }{}^{\mu \nu \alpha} \partial_\alpha \phi \Big).
    \end{aligned}
\end{equation}

The connection equation \eqref{Gfieldeqs} is not projectively coherent. In fact, it can be checked that the Palatini tensor \eqref{palatinite} is projectively invariant, while the hypermomentum \eqref{hypedef} is not, in general. The projective invariance of the gravity action imposes constraints on the matter sector. In particular, it requires a vanishing dilation current \cite{Hehl:1994ue,Vitagliano:2010sr,iosifidis2019exactly}, $\Delta^{\mu}:=\Delta^{\alpha\beta\mu}g_{\alpha\beta}\equiv 0$. In our current setting, such a constraint leads to the unphysical condition $\partial_{\mu}\phi=0$, implying that the scalar field cannot propagate. In order to avoid this issue, the matter sector must be separately projectively invariant on its own. We therefore impose projective coherence in the matter sector. This can be done by taking the trace of the connection field equations \eqref{Gfieldeqs} and requiring the resulting condition to hold identically. Doing so, we find the relation among the $\mathcal{C}_i$ functions:
\begin{equation}\label{constr}
  \mathcal{C}_1 = \frac{1}{16}(-4\, \mathcal{C}_2 +3\, \mathcal{C}_3).
\end{equation}
Imposing \eqref{constr} to the functions $\mathcal{C}_i$ from the beginning makes the connection equation \eqref{Gfieldeqs} projectively coherent. 

Having a Lagrangian that is invariant under projective transformations \eqref{projective}, we are free to fix one of the torsion or nonmetricity vectors by a projective gauge choice. We choose the gauge in which the $q^\mu$ nonmetricity vector vanishes, $q^\mu = 0$, and then solve the connection equation. Following the methods of \cite{iosifidis2019exactly, Iosifidis:2021ili}, the tensorial equation \eqref{Gfieldeqs} can be solved for the torsion and nonmetricity tensors. We obtain \cite{Andrei:inprep}
\begin{equation}\label{Q}
   Q_{\mu \lambda \nu } = \frac{5 \bigl(4 \mathcal{C}_2 + \mathcal{C}_3\bigr) }{4 \mathcal{A}} g_{\lambda \nu } \partial_{\mu } \phi-  \frac{\bigl(4 \mathcal{C}_2 + \mathcal{C}_3\bigr)}{4 \mathcal{A}}(g_{\mu \nu } \partial_{\lambda }\phi + g_{\lambda \mu } \partial_{\nu } \phi),
\end{equation}
\begin{equation}\label{S}
 S_{\mu \lambda \nu } = - \frac{\mathcal{C}_4 }{\mathcal{A}} \epsilon_{\mu \lambda \nu }{}^{\rho } \partial_{\rho }\phi -  \frac{ \bigl(6 \mathcal{C}_2 + \mathcal{C}_3 - \mathcal{A}^{\prime}\bigr)}{4 \mathcal{A}}(- g_{\mu \nu } \partial_{\lambda }\phi + g_{\lambda \nu } \partial_{\mu } \phi),
\end{equation}
where the projective invariance condition \eqref{constr} has already been taken into account. Substituting \eqref{Q} and \eqref{S} back into the action \eqref{action}, the theory can be reformulated as a metric theory, up to the boundary term displayed explicitly,
\beq
S[g,\phi]=\int d^{4}x \sqrt{-g}\qty[\frac{1}{2}\mathcal{A}(\phi)\tilde{R}- \frac{1}{2}\,\mathcal{K}(\phi) \partial_\mu \phi \partial^\mu \phi - \,\mathcal{V}(\phi) + \frac{1}{2}\tilde{\nabla}_\lambda \qty(\mathcal{A}(\phi)(- 4\, S^{\lambda} + q^{\lambda } - Q^{\lambda }))],
\eeq
where $\tilde R$ is the metric-dependent Ricci scalar constructed with the Levi-Civita connection while the kinetic function $\mathcal{K}(\phi)$ is given by
\begin{equation}\label{EkC}
\mathcal{K}(\phi) =  \mathcal{B} + \frac{9 \mathcal{C}_2^2}{2 \mathcal{A}} + \frac{9  \mathcal{C}_2 \mathcal{C}_3}{4 \mathcal{A}} -  \frac{3  \mathcal{C}_3^2}{32 \mathcal{A}} + \frac{6  \mathcal{C}_4^2}{\mathcal{A}} -  \frac{3 \mathcal{C}_3  \mathcal{A'}}{2 \mathcal{A}} -  \frac{3  \mathcal{A'}^2}{2 \mathcal{A}}.
\end{equation}
Notice that also in \eqref{EkC} the condition for projective invariance \eqref{constr} has already been implemented.

As a final step, we perform a conformal rescaling of the metric,
\begin{equation}
\label{eq: conformal transformation}
g^E_{\mu\nu} \equiv \frac{\mathcal{A}(\phi)}{M_P^2} g^J_{\mu\nu}, 
\end{equation}
where we denoted the Einstein-frame metric $g^E_{\mu\nu}$, and the Jordan-frame metric $g^J_{\mu\nu}$. After the transformation \eqref{eq: conformal transformation}, the equivalent Einstein-frame action reads:
\begin{equation}
S_E = \int d^4x \sqrt{-g_E} \qty( \frac{M_P^2}{2}R_E -\frac{1}{2} K(\phi) \,\partial^\mu\phi \,\partial_\mu \phi - U(\phi)),
\end{equation}
where the Einstein-frame kinetic term is
\begin{equation}\label{EFkinetic}
 K (\phi) = M_P^2\qty( \frac{ \mathcal{K}}{\mathcal{A}}+\frac{3}{2} \frac{{\mathcal{A}^{\prime}}^2}{\mathcal{A}^2}) = M^2_P \qty(\frac{\mathcal{B}}{\mathcal{A}} +\frac{1}{\mathcal{A}^2}\qty(\frac{9 \mathcal{C}_2^2}{2} + \frac{9  \mathcal{C}_2 \mathcal{C}_3}{4} -  \frac{3  \mathcal{C}_3^2}{32} + 6\, \mathcal{C}_4^2 -  \frac{3 \mathcal{C}_3  \mathcal{A'}}{2})),
\,
\end{equation}
and potential
\begin{equation}\label{potential}
  U(\phi)=\frac{M_P^4}{\mathcal{A}(\phi)^2}\mathcal{V}(\phi).
\end{equation}
The non-trivial kinetic factor \eqref{EFkinetic} gives rise to interesting consequences for inflationary phenomenology. To reveal those, we first define the canonical Einstein-frame scalar field:
\begin{equation}\label{eq:canonical_scalar_1}
\frac{d \chi}{d \phi} = \sqrt{M^2_P \qty(\frac{\mathcal{B}}{\mathcal{A}} +\frac{1}{\mathcal{A}^2}\qty(\frac{9 \mathcal{C}_2^2}{2} + \frac{9  \mathcal{C}_2 \mathcal{C}_3}{4} -  \frac{3  \mathcal{C}_3^2}{32} + 6\, \mathcal{C}_4^2 -  \frac{3 \mathcal{C}_3  \mathcal{A'}}{2}))}.   
\end{equation}
The Einstein-frame action in terms of the canonical scalar field then becomes:
\begin{equation}
S_E = \int d^4x \sqrt{-g_E} \qty( \frac{M_P^2}{2}R_E -\frac{1}{2} \,\partial^\mu\chi \,\partial_\mu \chi - U(\chi)).
\end{equation}
All the contributions coming from the original metric-affine theory are now encoded into the potential $U(\chi)$  embedded in the standard metric formulation, where the gravitational action is described by the usual Einstein-Hilbert term.

In order to make the model more predictive, we restrict the form of the coupling functions as follows
\begin{equation}\label{eq:couplings:general}
    \mathcal{A}'(\phi) = 2 \xi \mathcal{C}(\phi)\, ,
    \hspace{1 cm} 
    \mathcal{C}_i(\phi) = \xi_i \, \mathcal{C} (\phi) \, ,
\end{equation}
where $\xi, \xi_i$ are dimensionless couplings. Here the $\mathcal A(\phi)$ and the functions $\mathcal C_i(\phi)$ are controlled by the same scalar-field dependence. The scalar field redefinition \eqref{eq:canonical_scalar_1} reduces to
\begin{equation}\label{eq:canonical_scalar_general}
 \frac{d \chi}{d \phi} = M_P \sqrt{ \qty(\frac{\mathcal{B}}{\mathcal{A}} +\frac{\bar\xi \, \mathcal{C}^2}{\mathcal{A}^2})} \,, 
\end{equation}
where the effective coupling $\bar \xi$ is defined as
\begin{equation}\label{eq:barxi}
\bar \xi = -3 \xi  \xi_3+\frac{9 \xi_2 \xi_3}{4}+\frac{9 \xi_2^2}{2}-\frac{3 \xi_3^2}{32}+6 \xi_4^2 \,.
\end{equation}
From Eq.\ \eqref{eq:canonical_scalar_general}, we see that, if the inflationary dynamics is so that during slow-roll 
\begin{equation}
    \mathcal{A}(\phi) \sim M_P^2 \, , \qquad M_P^2 \mathcal{B}(\phi) \ll \bar\xi \, \mathcal{C}(\phi)^2 \label{eq:conditions}
\end{equation}
then we can approximate
\begin{equation}\label{eq:canonical_scalar_3}
 \frac{d \chi}{d \phi} \approx \sqrt{\bar\xi} \, \frac{\mathcal{C}(\phi)}{M_P}  = \sqrt{\bar\xi} \, \frac{\mathcal{I}'(\phi)}{M_P} \, ,
\end{equation}
where $\mathcal I'(\phi)=\mathcal C(\phi)$, so that $\mathcal I_i(\phi)=\xi_i\mathcal I(\phi)$ up to irrelevant integration constants. Then it is immediate to solve Eq.\ \eqref{eq:canonical_scalar_3} to obtain
\begin{equation}
    \chi(\phi) \approx \sqrt{\bar\xi} \, \frac{\mathcal{I}(\phi)}{M_P} \, .
\end{equation}
Therefore, if the conditions \eqref{eq:couplings:general} and \eqref{eq:conditions} are satisfied, the canonical scalar field redefinition is proportional to the non-minimal coupling functions $\mathcal {I}_i$. \footnote{For the purpose of proving this argument, here we also have assumed a positive $\bar \xi$ in \eqref{eq:conditions}. In the following, we will drop this assumption and consider the case $\bar\xi > 0$ and $\bar \xi < 0$, separately.} 
In the following sections, we study the implications of this setup for slow-roll inflation.

\section{Slow-roll inflation}\label{sec:sr}

Having formulated the theory in the Einstein frame, we can introduce the standard slow-roll formalism for the computation of the inflationary observables. Keeping the Planck mass explicit, we define the slow-roll parameters as
\begin{equation}\label{eq:sr}
    \epsilon_U (\chi) := \frac{M_P^2}{2}\qty(\frac{U'(\chi)}{U(\chi)})^2 ,\qquad \eta_U (\chi):= M_P^2\frac{U''(\chi)}{U(\chi)},
\end{equation}
where the prime $'$ denotes the derivative with respect to the canonical field $\chi$. Slow-roll inflation takes place while $\epsilon_U(\chi), \eta_U(\chi)\ll 1$ and it ends when either $\epsilon_U(\chi) = 1$ or $\eta_U(\chi) = 1$. The slow-roll parameters can be directly related to the CMB observables, in particular the tensor-to-scalar ratio $r$,  the spectral index $n_s$, the amplitude of the scalar perturbations $A_s$, and the number of $e$-folds $N_*$. In leading order, we have
\begin{align}\label{eq:sr_1}
r &= 16 \,\epsilon_U(\chi_*), \\ \label{eq:sr_2}
n_s &= 1 - 6\, \epsilon_U(\chi_*)+ 2 \,\eta_U (\chi_*),\\ \label{eq:sr_3}
A_s &= \frac{U(\chi_*)}{24\pi^2 M_P^4\,\epsilon_U(\chi_*)},\\ \label{eq:sr_4}
N_* &= \frac{1}{M_P^2} \int_{\chi_{\rm end}}^{\chi_*}\frac{U(\chi)}{U'(\chi)} d\chi,
\end{align}
where $\chi_*$ is the value of the field at the time at which the pivot scale $k_*$ exits the horizon during inflation and $\chi_{\rm end}$ denotes its value at the end of inflation\footnote{For the less common cases where $\eta_U$ reaches one before $\epsilon_U$, the end of inflation is instead defined by $\eta_U(\chi_{\rm end}) = 1$} i.e.\ such that $\epsilon_U(\chi_{\rm end})=1$. With these expressions, we can proceed to compute the predictions of the model.

\section{The model}\label{sec:model}

We now specialize the general ansatz \eqref{eq:couplings:general} by choosing
$\mathcal{C}(\phi)=\phi$ and fixing the integration constant in $\mathcal A$
to be $M_P^2$. Notice that in \eqref{eq:couplings:general} no assumption has been made on the magnitude of the $\xi, \bar \xi$ couplings, nor their sign yet. Together with the canonical choice $\mathcal B(\phi)=1$, this gives
\begin{equation}\label{eq:couplings}
\begin{aligned}
    &  \mathcal{A}(\phi) = {M_P^2} +\xi\, \phi^2  \,,\hspace{1 cm} 
    \mathcal{B}(\phi) = 1, 
    \,\hspace{1 cm} 
    \mathcal{C}_i(\phi) = \xi_{i} \,\phi \, .
    \end{aligned}
\end{equation}
From \eqref{EFkinetic} we obtain
\begin{equation}
 \begin{aligned}
K(\phi) =\frac{M_P^2}{M_P^2 + \,\xi \phi^2}+  \frac{M_P^2\bar\xi\phi^2}{\qty(M_P^2 + \,\xi \phi^2)^2} =\frac{M_P^{2} \left(M_P^{2} + \left(\xi + \bar\xi\right)\phi^{2}\right)}{ \left(M_P^{2} + \xi\phi^{2}\right)^{2}},
\end{aligned}
\end{equation}
with the combined Nieh-Yan-like coupling parameter $\bar\xi$ defined in Eq.\ \eqref{eq:barxi}. Let us note that for $\xi+\bar\xi \geq 0$ the kinetic function $K(\phi)$ is everywhere positive and the model is well behaved. Otherwise for $\xi+\bar\xi<0$, at large absolute values of the scalar field, $\phi^2>\frac{M_P^2}{|\xi +\bar\xi|}$, the kinetic function changes the sign and the model will become unstable and not relevant for our study.

\begin{figure}[t!]%
    \centering
    {\includegraphics[width=1\textwidth]{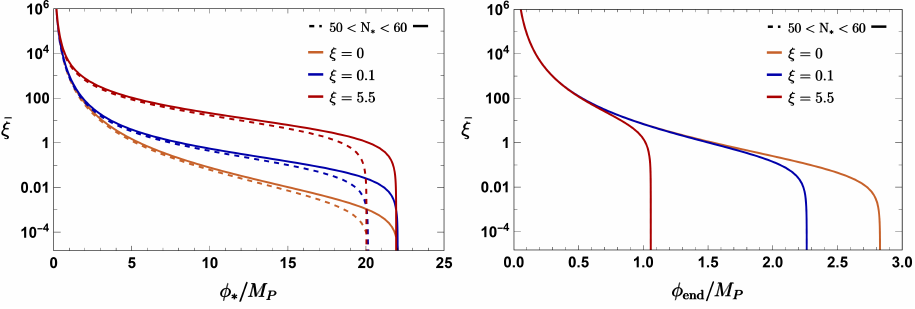}}%

   \caption{Relation between the non-minimal coupling $\bar \xi$ and the value of the scalar field at the pivot scale $\phi_*$ (left), relation between the non-minimal coupling $\bar \xi$ and the value of the scalar field at the end of inflation $\phi_{\rm end}$ (right). For each $\xi$ there is a value of $\bar \xi$ such that the field displacement turns sub-Planckian during inflation. The specific plots refer to the case considered in Fig.\ref{fig:model}, but the qualitative behavior is the same for any $\mathcal{V}(\phi) = \lambda \phi^k$. Notice that the scalar field always becomes sub-Planckian during inflation as $\bar \xi \rightarrow +\infty$.}
    \label{fig:xi_vs_phi}
\end{figure}

We then proceed by defining the canonical scalar field through \eqref{eq:canonical_scalar_1}
\begin{equation}\label{eq:canonical_scalar}
\frac{d \chi}{d \phi} = \sqrt{\frac{M_P^2}{M_P^2 + \,\xi \phi^2}+  \frac{M_P^2\bar\xi\phi^2}{\qty(M_P^2 + \,\xi \phi^2)^2}}= M_P\sqrt{\frac{M_P^2+(\xi+\bar \xi)\phi^2}{(M_P^2+\xi\phi^2)^2}}    .
\end{equation}
It is then clear that the argument of the square root is positive as long as $\xi+\bar \xi>0$.
This expression, in general, can only be integrated numerically, so an analytical expression for the inflationary observables in terms of the canonical Einstein-frame field $\chi$ cannot be obtained.
To compute the inflationary observables, we need to evaluate the derivatives of the potential $U(\phi(\chi))$ in terms of the canonical field $\chi$. This can be done without explicitly integrating \eqref{eq:canonical_scalar}, by using the chain rule in the computation. In the following, we consider separately $(\xi\geq0,\bar \xi\geq0)$ and  $(\xi\geq 0,\bar \xi\leq0)$, since they lead to different predictions. We do not consider the case  $(\xi< 0,\bar \xi>0)$ since $\xi<0$ induces a pole in the Einstein-frame potential, making the case not relevant for inflation.
\subsection{The regime $\xi\geq 0, \hspace{0.08cm} \bar{\xi}\geq 0$}
In the large $\bar \xi$ limit, simple analytical expressions can be obtained under the following considerations and assumptions.
We consider the parameter regime
\begin{equation}
 \xi\geq 0 ,\hspace{1cm} \bar \xi\geq 0,
\end{equation}
so that the kinetic function remains non-negative throughout the inflationary evolution. Notice from \eqref{eq:barxi} that the condition  $\bar \xi \geq 0$ is trivially satisfied if we assume $\xi_3 \rightarrow 0$. In general, however, we will need to satisfy the condition:
\begin{equation}
-16 \xi +12 \xi_2 -8 \sqrt{4 \xi ^2-6 \xi  \xi_2+3 \xi_2^2+\xi_4^2} < \xi_3 < -16 \xi +12 \xi_2 + 8 \sqrt{4 \xi ^2-6 \xi  \xi_2+3 \xi_2^2+\xi_4^2} \,.
\end{equation}
By looking at \eqref{eq:canonical_scalar} we can understand the effect of the non-minimal coupling $\bar \xi$. For $\bar \xi \rightarrow +\infty$ the r.h.s. of \eqref{eq:canonical_scalar} reduces to:
\begin{equation}\label{eq:canonical_scalar_2}
\frac{d \chi}{d \phi} \approx \sqrt{ \frac{M_P^2\bar\xi\phi^2}{\qty(M_P^2 + \,\xi \phi^2)^2}}.
\end{equation}
One could then naively think that the behavior of the canonical scalar $\chi$ dramatically changes whether $\xi = 0$ or $\xi \neq 0$. 
However, this is not necessarily the case during inflation. Let us first consider the case $\xi = 0$. In this case one can see that \eqref{eq:canonical_scalar_2} immediately reduces to:
\begin{equation}\label{eq:linear}
    \frac{d \chi}{d \phi} \approx \sqrt{\bar \xi}\frac{\phi}{M_P}.
\end{equation}
This has already been derived in \cite{Rasanen:2018ihz} for the standard non-minimally coupled Nieh-Yan term.
However, it can be shown numerically that this also happens for $\xi \neq 0$, in the case of a Jordan-frame potential in the monomial form
\begin{equation}\label{eq:monomial}
 \mathcal{V}(\phi) = \lambda M_P^{4}\qty(\frac{\phi}{M_P})^k.
\end{equation}
From Fig.\ref{fig:xi_vs_phi} we notice that $\bar \xi \gg 1$ implies $\phi_*/M_P\ll 1,\phi_{\rm end}/M_P \ll 1$ and the relation between the canonical scalar and the Jordan-frame scalar always reduces to \eqref{eq:linear} during inflation (this is equivalent to take  $\phi \rightarrow 0$ in \eqref{eq:canonical_scalar}). 
Hence, in the large $\bar \xi$ limit, the scalar field displacement is sub-Planckian during inflation and $\chi(\phi) \sim \phi^2$. 
By integrating \eqref{eq:linear} and plugging it into \eqref{eq:monomial} we find that
\begin{equation}\label{eq:Einstein_potential}
 U(\phi(\chi)) \sim 2^{k/2} \lambda \, \,\bar \xi^{-k/4} M_P^{4}\qty(\frac{\chi}{M_P})^{k/2},\,\,\,
\end{equation}
since $\mathcal A(\phi) \sim M_P^2$ in this limit. This means that if the Jordan-frame potential $\mathcal{V} \propto \phi^k$, then the Einstein-frame potential $U \sim \chi^{k/2}$ in the limit $\bar \xi \rightarrow \infty$. This is one of the central results of this work.
Moreover, in this limit we can explicitly write the predictions for the CMB observables by using \eqref{eq:sr_1}-\eqref{eq:sr_4}:
\begin{align}\label{eq:cmb_1}
r &\sim \frac{2k}{N_*}, \\
n_s &\sim 1-\frac{4+k}{4 N_*},\\
A_s &\sim \frac{2^{k/2}\bar \xi^{-k/4}\lambda N_* \left(k N_*\right)^{k/4}}{3\pi ^2 k},\\
N_* &\sim \frac{\chi_*^2}{k M_P^2},\label{eq:cmb_4}
\end{align}
where, in the last expression, we have neglected the contribution coming from the end of inflation $\chi_{\rm end}$.
In the following, we study numerically two examples of slow-roll inflation: quartic ($k=4$) and quadratic ($k=2$) Jordan-frame potentials. We show explicitly that in the large $\bar \xi$ coupling limit the slow-roll predictions correspond respectively to an effective quadratic and linear Einstein-frame potential. Moreover, we show that for intermediate values of $\bar \xi$, the new coupling can improve the predictions of Palatini inflation according to the most recent combination of dataset coming from CMB observations.

\subsubsection{Quartic potential}
\begin{figure}[t!]%
    \centering
    {\includegraphics[width=1\textwidth]{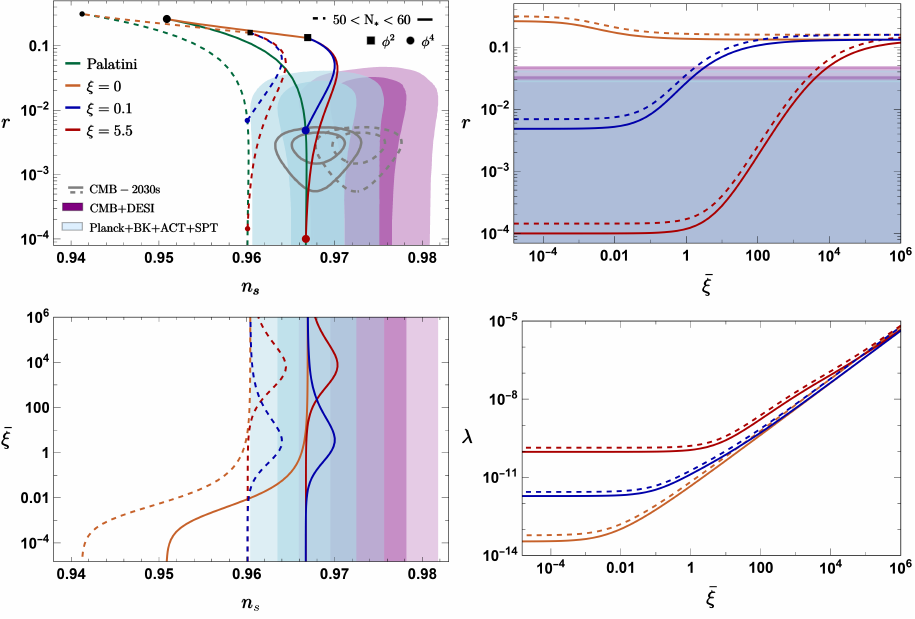}}%

   \caption{Slow roll predictions for the quartic Jordan-frame potential $\mathcal{V}(\phi) = \lambda \phi^4$. In panel (a) $r$ vs. $n_s$, in (b)  $r$ vs.  $\bar{\xi}$, (c) $n_s$ vs. $\bar{\xi}$, and (d) $\lambda$ vs.  $\bar{\xi}$ for $\xi =$ 0 (orange), 0.1 (blue), 5.5 (red). The green lines show the non-minimal Palatini model (i.e. $\bar \xi = 0$). The black dots represent the predictions for the quartic potential. The black squares represent the ones for the quadratic potential, and match the analytical predictions in \eqref{eq:cmb_1}-\eqref{eq:cmb_4} in the limit $\bar \xi \rightarrow \infty$.  All observables are computed at $N_* = 50$ (dashed lines), $N_* = 60$ (continuous lines). All the lines in the plot are computed numerically through Eqs. \eqref{eq:sr}-\eqref{eq:sr_4} by varying the parameters $\xi, \bar{\xi}$. The relation between $\lambda$ and $\bar \xi$ in (d) is obtained by fixing the amplitude of the power spectrum $A_s$ to its observed value $A_s \sim 2.1 \cdot 10^{-9}$. The cyan contours show the 95\% and 68\% confidence levels based on the latest combination of Planck, BICEP/Keck, ACT and SPT \cite{Balkenhol:2025wms}, the purple contours show the 95\% and 68\% confidence levels after including DESI to the aforementioned CMB observations, while the gray contours show the 95\% and 68\% confidence levels based on the projection of CMB observations during the next decade, using the Starobinsky model as prior. Details in the text.}
    \label{fig:model}
\end{figure}

Let us first consider the quartic Jordan-frame potential
\begin{equation}
    \mathcal{V}(\phi) = \lambda \phi^4.
\end{equation}
We show in Fig.~\ref{fig:model} (a) $r$ vs. $n_s$, (b)  $r$ vs.  $\bar{\xi}$, (c) $n_s$ vs. $\bar{\xi}$, and (d) $\lambda$ vs.  $\bar{\xi}$ for $\xi =$ 0 (orange), 0.1 (blue), 5.5 (red) for a quartic Jordan-frame potential $\mathcal{V}(\phi) = \lambda \phi^4$. The green lines show the non-minimal Palatini model (i.e. $\bar \xi = 0$).The black dots represent the predictions for the quartic potential. The black squares represent the ones for the quadratic potential, and match the analytical predictions in \eqref{eq:cmb_1}-\eqref{eq:cmb_4} in the limit $\bar \xi \rightarrow \infty$.  All observables are computed at $N_* = 50$ (dashed lines), $N_* = 60$ (continuous lines). All the lines in the plot are computed numerically through equations \eqref{eq:sr}-\eqref{eq:sr_4} by varying the parameters $\xi, \bar{\xi}$. The relation between $\lambda$ and $\bar \xi$ in (d) is obtained by fixing the amplitude of the power spectrum $A_s$ to its observed value $A_s \sim 2.1 \cdot 10^{-9}$. The cyan contours show the 95\% and 68\% confidence levels based on the latest combination of Planck, BICEP/Keck, ACT and SPT \cite{Balkenhol:2025wms}, the purple contours show the 95\% and 68\% confidence levels after including DESI to the aforementioned CMB observations, while the gray contours show the 95\% and 68\% confidence levels based on the projection of CMB observations during the next decade, using the Starobinsky model as prior.
For $\bar{\xi}=0$ we recover the extensively studied case of non-minimally coupled inflation in the Palatini formulation \cite{Jarv:2017azx, Racioppi:2018zoy, Racioppi:2019jsp, Gialamas:2020snr, Gialamas:2021rpr, Dioguardi:2025mpp}. 
The effect of increasing $\bar \xi$ can be understood from the orange, blue and red lines in Fig.\ref{fig:model}. For the case $\xi = \bar \xi = 0$ we recover the usual prediction for a quartic potential (black dots), as expected. Fixing $\xi = 0$, i.e. in the absence of a non-minimal coupling to the Ricci scalar, and turning on the Nieh-Yan-like coupling $\bar \xi$, the tensor-to-scalar ratio $r$ is suppressed and the spectral index $n_s$ increases until $\bar \xi \sim 10^2$ (orange lines). For values $\bar \xi \gtrsim 10^2$ the predictions correspond to those of quadratic inflation (black squares), as already shown in \cite{Langvik:2020nrs} for the standard Nieh-Yan coupling. For non-zero $\xi$, increasing the Nieh-Yan-like coupling $\bar \xi$ (blue and red lines) induces a larger tensor-to-scalar ratio $r$, (see panel (b) of Fig.\ref{fig:model}), as well as a larger spectral index $n_s$ (see panel (c) of Fig.\ref{fig:model}). Eventually, for sufficiently large values of $\bar \xi$ the model approaches the quadratic inflation predictions for $r, n_s$, as expected from \eqref{eq:Einstein_potential}. The value $\bar \xi$ at which this asymptotic regime is reached depends on the value $\xi$ of the non-minimal Ricci coupling. For the parameter choices shown in Fig.\ref{fig:model}, this happens for $\bar \xi \sim 10^3$ if $\xi = 0.1$ and for $\bar \xi \sim 10^6$ if $\xi = 5.5$. This feature is particularly interesting in light of current and future CMB observations. The slight increase in $n_s$ produced by the Nieh-Yan-like coupling can restore the $2\sigma$ compatibility of the Palatini non-minimally coupled model for $N_*=50$ $e$-folds, even for $\bar \xi<1$. Moreover, even in the case of a very large non-minimal Ricci coupling $\xi$, the model can predict a value of the tensor-to-scalar ratio $r$ that is in the range of next-generation CMB observations, providing a testable deviation from the standard Palatini prediction.
\subsubsection{Quadratic potential}
\begin{figure}[t!]%
    \centering
    {\includegraphics[width=1\textwidth]{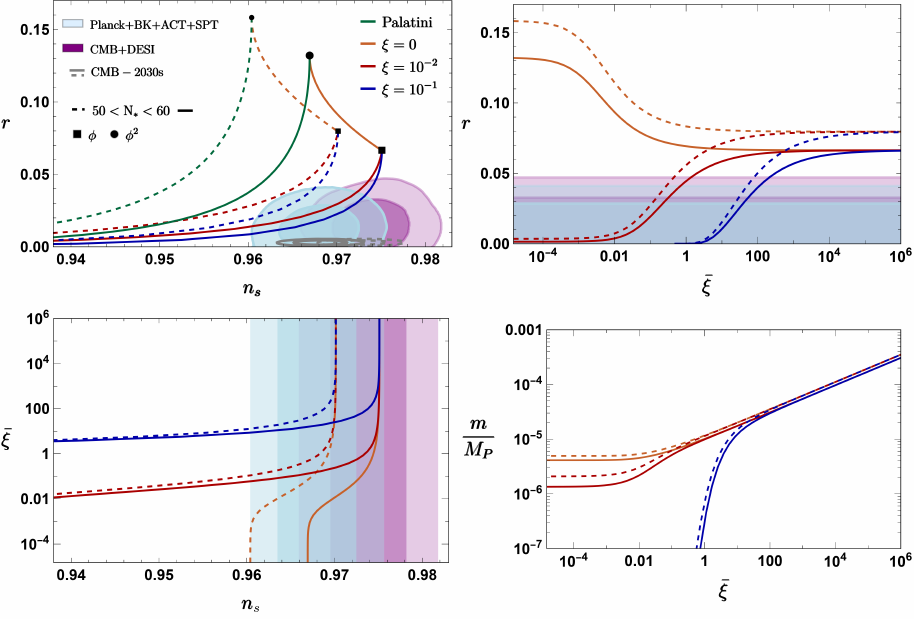}}%

   \caption{Slow roll predictions for the quadratic Jordan-frame potential $\mathcal{V}(\phi) = m^2 \phi^2$. In panel (a) $r$ vs. $n_s$, in (b)  $r$ vs.  $\bar{\xi}$, in (c) $n_s$ vs. $\bar{\xi}$, and in (d) $m/M_P$ vs. $\bar{\xi}$ for $\xi =$ 0 (orange), $10^{-2}$ (red), $10^{-1}$ (blue).  The green lines show the non-minimal Palatini model (i.e. $\bar \xi = 0$). The black dots represent the predictions for the quadratic potential. The black squares represent the ones for the linear potential, and match the analytical predictions in \eqref{eq:cmb_1}-\eqref{eq:cmb_4} in the limit $\bar \xi \rightarrow \infty$. All observables are computed at $N_*=50$ (dashed lines), $N_* = 60$ (continuous lines). All the lines in the plot are computed numerically through Eqs. \eqref{eq:sr}-\eqref{eq:sr_4} by varying the parameters $\xi, \bar{\xi}$. The relation between $m/M_P$ and $\bar \xi$ in (d) is obtained by fixing the amplitude of the power spectrum $A_s$ to its observed value $A_s \sim 2.1 \cdot 10^{-9}$. The cyan contours show the 95\% and 68\% confidence levels based on the latest combination of Planck, BICEP/Keck, ACT and SPT \cite{Balkenhol:2025wms},  the purple contours show the 95\% and 68\% confidence levels after including DESI to the aforementioned CMB observations, while the gray contours show the 95\% and 68\% confidence levels based on the projection of CMB observations during the next decade, by using the Starobinsky model as prior. Details in the text.}
    \label{fig:model_2}
\end{figure}

Let us now consider the case of a quadratic Jordan-frame potential
\begin{equation}
    \mathcal{V}(\phi) = m^2\phi^2,
\end{equation}
where we identify $m^2=\lambda M_P^2$ in \eqref{eq:monomial}.
We show in Fig.\ref{fig:model_2} (a) $r$ vs. $n_s$, (b)  $r$ vs.  $\bar{\xi}$, (c) $n_s$ vs. $\bar{\xi}$, and (d) $m/M_P$ vs. $\bar{\xi}$ for $\xi =$ 0 (orange), $10^{-2}$ (red), $10^{-1}$ (blue) for a quadratic Jordan-frame potential $\mathcal{V}(\phi) = m^2 \phi^2$.  The green lines show the non-minimal Palatini model (i.e. $\bar \xi = 0$). The black dots represent the predictions for the quadratic potential. The black squares represent the ones for the linear potential, and match the analytical predictions in \eqref{eq:cmb_1}-\eqref{eq:cmb_4} in the limit $\bar \xi \rightarrow \infty$. All observables are computed at $N_*=50$ (dashed lines), $N_* = 60$ (continuous lines). All the lines in the plot are computed numerically through Eqs. \eqref{eq:sr}-\eqref{eq:sr_4} by varying the parameters $\xi, \bar{\xi}$. The relation between $m/M_P$ and $\bar \xi$ in (d) is obtained by fixing the amplitude of the power spectrum $A_s$ to its observed value $A_s \sim 2.1 \cdot 10^{-9}$. The cyan contours show the 95\% and 68\% confidence levels based on the latest combination of Planck, BICEP/Keck, ACT and SPT \cite{Balkenhol:2025wms},  the purple contours show the 95\% and 68\% confidence levels after including DESI to the aforementioned CMB observations, while the gray contours show the 95\% and 68\% confidence levels based on the projection of CMB observations during the next decade, using the Starobinsky model as prior. When the Ricci scalar is non-minimally coupled with inflaton through a non-minimal coupling $\mathcal{A}(\phi)$ as in \eqref{eq:couplings}, the resulting Einstein-frame potential suffers from the so-called $\eta$-problem for values of the coupling $\xi\gtrsim 10^{-2}$. This is a well-known aspect of Palatini inflation \cite{Tenkanen:2017jih}. As we can see from Fig.\ref{fig:model_2}, introducing a non-minimal coupling $\bar \xi$ solves the issue. As expected from \eqref{eq:Einstein_potential}, the predictions reduce (in the large $\bar \xi$ coupling limit) to those of linear inflation. In particular, for values of the coupling $10^{-2}\lesssim\bar \xi\lesssim 10^2$, the $\bar \xi$ coupling shifts the slow-roll predictions into the region compatible with current observations, increasing both the tensor-to-scalar ratio $r$ and the spectral index $n_s$. This makes the model testable and falsifiable with next generation probes. We recall that also in this case the Jordan-frame field moves over sub-Planckian values during inflation, for sufficiently large $\bar \xi$.

\subsection{The regime $\xi\geq 0,\hspace{0.08cm} \bar \xi\leq 0$}

\begin{figure}[t!]%
    \centering
    {\includegraphics[width=1\textwidth]{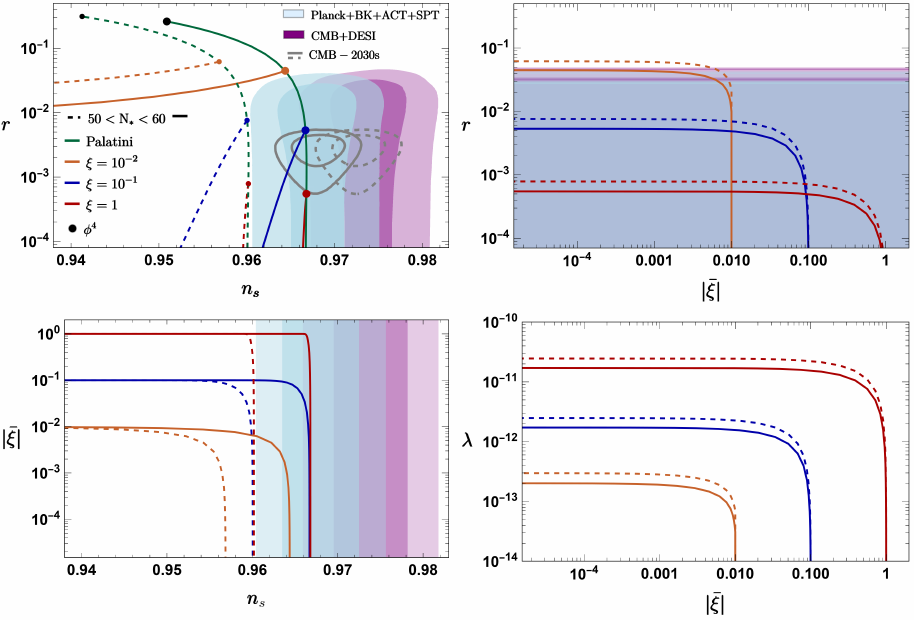}}%

   \caption{Slow-roll predictions for the quartic Jordan-frame potential $\mathcal{V}(\phi)=\lambda\phi^4$ in the negative-$\bar\xi$ regime. The various panels show the following. In panel (a) $r$ vs. $n_s$, in (b)  $r$ vs.  $|\bar{\xi}|$,in (c) $n_s$ vs. $|\bar{\xi}|$, and in (d) $\lambda$ vs.  $|\bar{\xi}|$ for $\xi = 10^{-2}$ (orange), $10^{-1}$ (blue), $1$ (red) for a quartic Jordan-frame potential $\mathcal{V}(\phi) = \lambda \phi^4$. The green lines show the non-minimal Palatini model (i.e. $\bar \xi = 0$). The black dots represent the predictions for the quartic potential. The predictions \eqref{eq:cmb_neg_1}-\eqref{eq:cmb_neg_4}, obtained in the analytical limit $ |\bar\xi| \rightarrow \xi$, are not visible in the plots because they lie outside the plot range, for the chosen values of $\xi$ (see in particular eq.\eqref{eq:cmb_neg_2}).  All observables are computed at $N_*=50$ (dashed lines), $N_* = 60$ (continuous lines). All the lines in the plot are computed numerically through Eqs. \eqref{eq:sr}-\eqref{eq:sr_4} by varying the parameters $\xi, |\bar{\xi}|$. The relation between $\lambda$ and $|\bar \xi|$ in (d) is obtained by fixing the amplitude of the power spectrum $A_s$ to its observed value $A_s \sim 2.1 \cdot 10^{-9}$. The cyan contours show the 95\% and 68\% confidence levels based on the latest combination of Planck, BICEP/Keck, ACT and SPT \cite{Balkenhol:2025wms},  the purple contours show the 95\% and 68\% confidence levels after including DESI to the aforementioned CMB observations, while the gray contours show the 95\% and 68\% confidence levels based on the projection of CMB observations during the next decade, by using the Starobinsky model as prior. Details in the text.}
    \label{fig:model_3}
\end{figure}

For completeness, we now consider the case in which the Ricci coupling is positive $\xi\geq 0$ but the Nieh-Yan-like one is negative $\bar \xi \leq 0$. We recall that in order to have a well-defined kinetic term we need to satisfy the condition $\xi+\bar \xi\geq 0$. In the limit $|\bar \xi|\ll \xi$ we have:
\begin{equation}
    \frac{d \chi}{d \phi} \approx \sqrt{\frac{M_P^2}{M_P^2+\xi \phi^2}},
\end{equation}
and the predictions reduce to those of Palatini inflation. When instead $|\bar \xi|\rightarrow\xi$ we get
\begin{equation}
    \frac{d \chi}{d \phi} \approx \frac{M_P^2}{M_P^2+\xi \phi^2},
\end{equation}
which can be directly integrated, giving:
\begin{equation}\label{eq:chi_neg}
   \chi \approx \frac{M_P}{\sqrt{\xi}}\arctan(\frac{\sqrt{\xi} \phi}{M_P}).
\end{equation}
For a monomial Jordan-frame potential of the form \eqref{eq:monomial}, Eq. \eqref{eq:chi_neg} can be inverted in the range $-\frac{\pi}{2}<\sqrt{\xi}\,\chi/M_P<\frac{\pi}{2}$. This gives
\begin{equation}
  U(\chi) \approx  \frac{\lambda M_P^4 \xi ^{-k/2} \tan ^k\left(\frac{\sqrt\xi \chi}{M_P} \right)}{\left(1 + \tan ^2\left(\frac{\sqrt{\xi}\chi}{M_P} \right)\right)^2}.
\end{equation}
In this limit we can get (rather cumbersome) analytical expressions for the slow-roll observables:
\begin{align}\label{eq:cmb_neg}
    r &\sim \frac{128 k \xi}{(4-k) e^{16 \xi N_*}+2 (k-2) e^{8 \xi N_*}-k},\\
    n_s &\sim 1-\frac{16 \xi \left((k-4) e^{16 \xi N_*}-(k-2) e^{8 \xi N_*}-k\right)}{\left(e^{8 \xi N_*}-1\right) \left((k-4) e^{8 \xi N_*}-k\right)}, \\
    A_s &\sim \frac{\lambda  e^{-16 \xi N_*}}{3072 \pi ^2 k \xi} \left(e^{8 \xi N_*}-1\right) \left((4-k) e^{8 \xi N_*}+k\right)^3 \left(\frac{\sqrt{k-k e^{8 \xi N_*}}}{\sqrt{\xi \left((k-4) e^{8 \xi N_*}-k\right)}}\right)^k,\\
    N_* &\sim \frac{1}{8\xi}\ln \left(\tan ^2\left(\frac{\sqrt{\xi}\chi_*}{M_P}  \right)+1\right)-\frac{1}{8\xi}\ln \left((k-4) \tan ^2\left(\frac{\sqrt{\xi}\chi_*}{M_P}  \right)+k\right)+ \frac{\ln(k)}{8\xi}.
\end{align}
where again we have neglected the contribution coming from the end of inflation $\chi_{\rm end}$.
As an example, we now show the behavior of this model for the specific case of a quartic Jordan-frame potential $k=4$. In this case Eqs.\eqref{eq:cmb_neg} reduce to:
\begin{align}\label{eq:cmb_neg_1}
    r &\sim \frac{128\xi}{e^{8\xi N_*}-1},\\
    n_s &\sim 1-8\xi-\frac{3}{16}r,\label{eq:cmb_neg_2}\\
    A_s &\sim \frac{\lambda  e^{-16 \xi N_*} \left(e^{8 \xi N_*}-1\right)^3}{192 \pi ^2 \xi^3},\\
    N_* &\sim \frac{\ln \left(\tan ^2\left(\frac{\sqrt{\xi}\, \chi_*}{M_P}\right)+1\right)}{8 \xi}.\label{eq:cmb_neg_4}
\end{align}
Fig.\ref{fig:model_3} shows the numerical results for the case $k=4$: in panel (a) $r$ vs. $n_s$, (b)  $r$ vs.  $|\bar{\xi}|$, (c) $n_s$ vs. $|\bar{\xi}|$, and (d) $\lambda$ vs.  $|\bar{\xi}|$ for $\xi = 10^{-2}$ (orange), $10^{-1}$ (blue), $1$ (red) for a quartic Jordan-frame potential $\mathcal{V}(\phi) = \lambda \phi^4$. The green lines show the non-minimal Palatini model (i.e. $\bar \xi = 0$). The black dots represent the predictions for the quartic potential. All observables are computed at $N_*=50$ (dashed lines), $N_* = 60$ (continuous lines). All the lines in the plot are computed numerically through Eqs. \eqref{eq:sr}-\eqref{eq:sr_4} by varying the parameters $\xi, |\bar{\xi}|$. The predictions \eqref{eq:cmb_neg_1}-\eqref{eq:cmb_neg_4}, obtained in the analytical limit $|\bar \xi| \rightarrow \xi$, are not visible in the plots because they lie outside the plot range, for the chosen values of $\xi$ (see in particular eq.\eqref{eq:cmb_neg_2}). The relation between $\lambda$ and $|\bar \xi|$ in (d) is obtained by fixing the amplitude of the power spectrum $A_s$ to its observed value $A_s \sim 2.1 \cdot 10^{-9}$. The cyan contours show the 95\% and 68\% confidence levels based on the latest combination of Planck, BICEP/Keck, ACT and SPT \cite{Balkenhol:2025wms},  the purple contours show the 95\% and 68\% confidence levels after including DESI to the aforementioned CMB observations, while the gray contours show the 95\% and 68\% confidence levels based on the projection of CMB observations during the next decade, by using the Starobinsky model as prior. For $\xi \gtrsim 5\cdot 10^{-3}$, the predicted value of $n_s$ becomes smaller than in the standard Palatini inflation, driving the prediction outside the data-compatible region. For $\xi \lesssim 5\cdot 10^{-3}$, the effect of $\xi$ becomes progressively smaller, and the predictions approach those of the minimally coupled quartic potential. We therefore conclude that the negative $\bar\xi$ regime does not improve the predictions of standard Palatini inflation. However, the model can still give distinct compatible predictions for values of $\xi$ where Palatini inflation is already compatible with data, and testable with the upcoming CMB observations.

\section{Conclusions}\label{sec:conclusion}
In this paper, we considered single-field slow-roll inflation in a metric-affine setting, with a scalar field non-minimally coupled to the non-Riemannian Ricci scalar, and derivative couplings between the scalar field and the torsion and nonmetricity vectors. 
We showed that, in the limit of a large positive coupling $\bar \xi$, the Einstein-frame canonical field $\chi$ and the Jordan-frame field $\phi$ are related by $\chi \sim \phi^2$. In the case of a monomial Jordan-frame potential $\mathcal V \propto \phi^k$, this implies an Einstein-frame potential $U \sim \chi^{k/2}$. We explicitly studied the numerical predictions for a quartic and quadratic Jordan-frame potential and showed that for intermediate values of $\bar \xi$ (with the precise interval depending on the potential) the coupling can restore compatibility of the Palatini model with cosmological observations. In particular, for the quartic potential, the model can predict values of the tensor-to-scalar ratio $r$ which fall in the window of next generation CMB observations for $\bar \xi\lesssim 10^4$, providing a testable and falsifiable prediction for slow-roll inflation in this setting. For the quadratic potential, the effective coupling $\bar \xi$ cures the $\eta$-problem that arises in the Palatini case for $\xi \gtrsim 10^{-2}$, and provides predictions compatible with the current observations for $10^{-2}\lesssim\bar\xi\lesssim 10^2$, testable with the upcoming CMB measurements, during the next decade. We also considered, for completeness, the regime $\bar\xi<0$ and found that, in general, the model does not improve the predictions of non-minimally coupled Palatini inflation.

A general theoretical treatment of non-minimally coupled scalar fields in metric-affine gravity, including the full structure of the metric-affine field equations and their metric-equivalent formulation, will be presented in a separate work in preparation~\cite{Andrei:inprep}.

The results in the present paper show that derivative couplings to the non-Riemannian sector can leave observable imprints even after the independent connection is integrated out, making metric-affine inflation with non-minimal couplings a testable extension of the Palatini framework.


\bigskip 
\begin{acknowledgments}
    This work was supported by the Estonian Research Council via the grants PRG1677, PRG2608, TARISTU24-TK10, TARISTU24-TK3, and the Centre of Excellence program ``Foundations of the Universe'' (TK202U1, TK202U4). This article is based upon work from COST Actions COSMIC WISPers CA21106 and CosmoVerse CA21136, supported by COST (European Cooperation in Science and Technology).
    DI's work was supported by the  Istituto Nazionale di Fisica Nucleare (INFN), Sezioni  di Napoli,  {\it Iniziative Specifiche} SKY.
\end{acknowledgments}

\bibliographystyle{utphys}
\bibliography{ref}

\end{document}